\documentclass[prb,
twocolumn,
superscriptaddress,showpacs,amsmath,amssymb]{revtex4}

\begin{document}

\author{G.E.~Volovik}
\affiliation{Low Temperature Laboratory, Aalto University,  P.O. Box 15100, FI-00076 Aalto, Finland}
\affiliation{Landau Institute for Theoretical Physics, acad. Semyonov av., 1a, 142432,
Chernogolovka, Russia}

\title{On the global temperature of the Schwarzschild-de Sitter spacetime}

\date{\today}

\begin{abstract}
It is shown that the Schwarzschild-de Sitter (SdS) spacetime has the universal temperature. This temperature describes the thermal processes of decay of the composite particles and the other processes, which are energetically forbidden in the Minkowski spacetime, but are allowed in the de Sitter and in SdS backgrounds. In particular, this temperature describes the probability of ionization of the atom in the SdS, which is observed by the stationary observer at the point where the shift function (velocity) in the Arnowitt-Deser-Misner formalism changes sign. This activation temperature does not depend on the black hole mass and is fully determined by the Hubble parameter, $T_{\rm SdS}=\sqrt{3}H/\pi$. This temperature is twice the Bousso-Hawking temperature $T_{\rm BH}$, which characterizes the limit of degenerate Lorentzian Schwarzschild-de Sitter universe, when the cosmological and black hole horizons are close to each other, $T_{\rm SdS}=2T_{\rm BH}$. 

The similar doubling of the temperature of Hawking radiation is known in the pure de Sitter spacetime, where the corresponding local temperature describing the ionization of atoms is twice the Gibbons-Hawking temperature, $T_{\rm dS}=2T_{\rm GH}=H/\pi$. We suggest that the activation temperature $T_{\rm dS}$ can be considered as the thermodynamic temperature of the de Sitter state, which determines the local entropy in this state, $s=3H/4G$.
\end{abstract}

\maketitle


\section{Introduction}

 The issue of the stability of the de Sitter vacuum is still an unsolved problem, see e.g.\cite{Kamenshchik2022,Polyakov2012}  and references therein. In spite of the Gibbons-Hawking radiation from the cosmological horizon, we do not know whether or not this leads to the decay of the de Sitter vacuum. 
It is not excluded that even if the particle creation by Gibbons-Hawking radiation takes place, the de Sitter expansion may immediately dilute the produced particles, and this prevents the vacuum decay in de Sitter spacetime.

 That is why one may think that study of the black hole in the de Sitter environment --  the Schwarzschild-de Sitter (SdS) spacetime with the presence of two horizons simultaneously -- can be even more difficult task. Each horizon is characterized by its own temperature, see e.g. Refs.\cite{BoussoHawking1996,Shankaranarayanan2003,Padmanabhan2007,Traschen2020}. The  Reissner-Nordstr\"om black hole also contains two horizons, but  the global Hawking temperature of radiation does exist.  It is determined by the correlations between the inner and outer horizons.\cite{Volovik2021,Volovik2022,Singha2022}   But the SdS has a different configuration of the two horizons, and it is not clear, whether such configuration allows the existence of some kind of the global temperature.\cite{Akhmedov2022} 

Here we consider the existence of the global temperature in dS and in SdS, which are not directly related to the horizons and to the corresponding Hawking radiation.  The temperature
$T_{\rm dS}$  and $T_{\rm SdS}$ describes the thermal activation processes correspondingly in the dS and SdS environments.  It is obtained when one considers for example the process of ionization of atoms in the dS and in the SdS spacetimes. This temperature describes also the other activation processes, such as the splitting of the composite particle, which is not possible in Minkowski spacetime. 

For the SdS spacetime, we find that in spite of existence of two horizons, the temperature $T_{\rm SdS}$ does not depend on the mass of the black hole, so it is really the global temperature. It is solely determined by the Hubble parameter $H$, but is by factor $2\sqrt{3}$ larger than the Gibbons-Hawking temperature $T_{\rm GH}=H/2\pi$ of Hawking radiation from the cosmological horizon in the de Sitter spacetime, $T_{\rm SdS}=2\sqrt{3}T_{\rm GH}$.

We use the extension of Painlev\'e-Gullstrand  coordinates, which describes the metric in the whole range of radial coordinates $0<r < \infty$.\cite{Volovik2023}  In this coordinate system there is the point (surface) $r=r_0$ at which the shift function (velocity) changes sign. At this point the observer is at rest, while the observers at $r<r_0$ are free falling to the black hole  and the observers at $r>r_0$  are free falling towards the cosmological horizon, see also Ref.\cite{ToporenskyZaslavskii2023}. 
Here is the main difference from the Reissner-Nordstr\"om black hole, where the same observer crosses two horizons, outer and inner, which allows to determine the single temperature of the Hawking radiation from the outer horizon of black hole.  

In the SdS spacetime, the temperatures of Hawking radiation from cosmological and black hole horizons and also the activation temperature are measured by the stationary observer at $r=r_0$. It was found by Bousso and Hawking\cite{BoussoHawking1996}  that the radiation temperatures differ by the normalization factor from the traditional Hawking temperatures determined by gravity at the horizons. These renormalized temperatures approach the same constant value, $T_b\rightarrow T_{\rm BH}-0$ and $T_c\rightarrow T_{\rm BH}+0$, in the Nariai limit when the cosmological and black hole horizons approach each other. 

It appears that there is the close connection between the global temperature $T_{\rm SdS}$ and the Bousso-Hawking temperature $T_{\rm BH}$. The global temperature is twice larger than the  Bousso-Hawking temperature in the Narai limit,  $T_{\rm SdS}=2T_{\rm BH}$. This is similar to the factor 2 in the de Sitter spacetime, where the activation temperature is twice larger than the Gibbons-Hawking temperature, $T_{\rm dS}=2T_{\rm GH}$. We discuss the possible thermodynamic origin of the factor 2 in the dS state.

\section{Modification  of Painlev\'e-Gullstrand  coordinates for Schwarzschild-de Sitter spacetime}

The relevant modification of  Painlev\'e-Gullstrand \cite{Painleve,Gullstrand} coordinates for Schwarzschild-de Sitter spacetime,  which describes the metric in the whole range of radial coordinates $0<r < \infty$,  is ($c=\hbar=1$):\cite{Volovik2023} 
\begin{equation}
ds^2 =g_{\mu\nu}dx^\mu dx^\nu=- N^2dt^2 +\frac{1}{N^2}(dr -v dt)^2+ r^2 d\Omega^2\,.
\label{PdGspherical}
\end{equation}
The lapse function $N$ and shift function  $v$ in the Arnowitt-Deser-Misner (ADM)  formalism\cite{ADM2008}  are given by:
\begin{equation}
N^2=1-C\,\,, \,\,    v^2(r)=(1-C)\left( \frac{2GM}{r}  +H^2r^2 -C\right)  \,.
\label{SdSADM}
\end{equation}
Here $C$ is the constant parameter, which is chosen in such a way, that the square of the shift velocity $v^2(r)$ has minimum at the point, where it is zero, $v(r_0)=0$.
This gives the following values of the parameter $C$ and of the stationary point $r_0$, where the velocity is zero:
\begin{equation}
C=3(GMH)^{2/3}  \,\,, \,\, r_0^3=\frac{GM}{H^2}  \,.
\label{parameters}
\end{equation}
The spherical surface at  $r=r_0$ is the zero-gravity surface.
Similar smooth  coordinate  frame expanding for large $r>r_0$ and
contracting for small $r<r_0$ was introduced in Ref.\cite{ToporenskyZaslavskii2023}, where in their notations $N=e_0$.

The square of shift velocity can be rewritten as
\begin{eqnarray}
  v^2(r)=(1-C)\left(\frac{2GM}{r}  +H^2r^2 -3(GMH)^{2/3}\right) =
\label{shiftSdSADM0}
  \\
=  \frac{C(1-C)}{3} \left(\frac{2r_0}{r} + \frac{r^2}{r_0^2} - 3\right)=
\label{shiftSdSADM1}
\\
= C(1-C) (r-r_0)^2 \frac{r+2r_0}{3rr_0^2}.
\label{shiftSdSADM2}
\end{eqnarray}
Then the shift velocity can be written in the form, which changes sign at $r=r_0$ as it should be for the free falling observer:
\begin{equation}
  v(r)= \sqrt{C(1-C)} \,\sqrt{\frac{r+2r_0}{3rr_0^2}}\,\,(r-r_0)\,.
\label{vSdSADM3}
\end{equation}
The observers at $r<r_0$ have velocity $v(r)<0$ and are free falling to the black hole, while the observers at $r>r_0$ have $v(r)>0$ and are free falling towards the cosmological horizon.
Near the stationary point  $r=r_0$ the shift velocity is:
\begin{equation}
  v(r)\approx \sqrt{3(1-C)}\,H(r-r_0) \,\,,\, |r-r_0|\ll r_0\,.
\label{vSdSADM4}
\end{equation}

\section{Hawking radiation in Schwarzschild-de Sitter spacetime}

The Hawking radiation from the two horizon is also modified, since the stationary observer is not at infinity in case of the black hole radiation and is not at $r=0$ in case of the de Sitter radiation. The static observer is at $r=r_0$, and that is why the frequency measured by this static observer  is red-shifted by the factor $N=(1-C)^{1/2}=1/\gamma_t$, where $\gamma_t$ is the normalization constant introduced in Ref. \cite{BoussoHawking1996}. As a result, the Hawking temperatures of black hole and cosmological horizons are renormalized:\cite{BoussoHawking1996}
\begin{equation}
T_b=\frac{T_{b0}}{N}\,\,,\,\, T_c=\frac{T_{c0}}{N} \,.
\label{SdSHawking}
\end{equation}
Here $T_{b0}=\kappa_b/2\pi$ and $T_{c0}=\kappa_c/2\pi$ are the original nonrenormalized Hawking temperatures of black hole and cosmological horizons, which are determined by gravity at the horizons. 

There is the special limit case, in which the global temperature of Hawking radiation can be determined for the SdS system. This is when the black hole horizon approaches the cosmological horizon, $r_b \rightarrow r_0-0$ and  $r_c \rightarrow r_0 + 0$. This takes place in the limit when $C\rightarrow 1$ and both temperatures approach the Bousso-Hawking value:\cite{BoussoHawking1996} 
\begin{equation}
T_{\rm BH}=T_b=T_c=\sqrt{3}\frac{H}{2\pi}=\frac{1}{6\pi GM}\,.
\label{GlobalT}
\end{equation}
The divergence in Eq.(\ref{SdSHawking}) due to $N\rightarrow 0$ is compensated by nullification of gravity in the vicinity of the stationary point.

For the general case, when $r_b < r_c$ and there are two Hawking temperatures, the existence of the thermal equilibrium in the SdS  spacetime is still an open problem. In our approach, the $T_{\rm SdS}$ is the temperature, which characterizes the processes of thermal activation, observed by the stationary observer at $r=r_0$. We consider these processes on example of the processes of ionization of an atom and decay of composite particles. These processes are not possible in Minkowski vacuum, where the nonzero temperature is required for activation. But they are possible in the dS spacetime, where they have been considered in Ref. \cite{Volovik2009}, and also in the SdS spacetime.

 \section{Particle spectrum and trajectories}

We study the activation processes using the semiclassical tunneling method, which uses the trajectories of particles.
 The energy spectrum of massive particles in the extended PG metric of SdS spacetime is given by the contravariant components of the metric:
\begin{eqnarray}
g^{\mu\nu}p_\mu p_\nu +m^2=0\,,
\label{spectrum1}
\\
-\frac{1}{N^2}(p_0 -vp_r)^2 +N^2 p_r^2  +p_\perp^2+ m^2=0 \,.
\label{spectrum2}
\end{eqnarray}
This gives for example the radial trajectories of particles $p_r(r,E)$, where $E=p_0$ is the fixed energy of the travelling particle. In particular, the radial trajectory of massless particle, $m=0$, is:
\begin{equation}
p_r(r,E)= -\frac{E}{N^2-v(r) }\,.
\label{pSdS}
\end{equation}
Near the poles it is expressed in terms of the static metric:
\begin{eqnarray}
p_r(r,E) \approx - \frac{2E}{f(r)}\,,
\label{NearPoles}
\\
f(r) = 1 - \frac{2MG}{r} - H^2 r^2\,.
\label{NearPoles2}
\end{eqnarray}

 \section{Probability of ionization and global temperature}
 
Let us now consider the rate of ionization of an atom as viewed by the static observer at $r=r_0$. We follow the tunneling approach\cite{Wilczek2000,Srinivasan1999,Volovik1999}, which has been applied  to the ionization in the dS backgound in Ref. \cite{Volovik2009}. We start with the electron at some energy level in the atom, which is at rest  at $r=r_0$. This corresponds to the spectrum in Eq.(\ref{spectrum2}) with $v=p_r=p_\perp=0$ and mass $m-\epsilon_0$, where $m$ is the mass of electron and $\epsilon_0 \ll m$ is the ionization potential:
\begin{eqnarray}
-\frac{p_0^2}{N^2}+ (m-\epsilon_0)^2=0 \,.
\label{Atom}
\end{eqnarray}
The free electron, which left the atom, obeys Eq.(\ref{spectrum2}) with $p_0/N=m-\epsilon_0$ from Eq.(\ref{Atom}):
\begin{eqnarray}
-\frac{1}{N^2}(N(m-\epsilon_0) -vp_r)^2 +N^2 p_r^2  + m^2=0 \,.
\label{Electron}
\end{eqnarray} 

First let us consider the semiclassical tunneling trajectory in the simplest case of small $C\ll 1$ (or $N\rightarrow 1$), when $T_c \ll T_b$. Then one obtains the following imaginary momentum on the tunneling trajectory:
\begin{equation}
{\rm Im} \,p_r(r)=\sqrt{2m\epsilon_0 -m^2 v^2(r)} \,.
\label{trajectory}
\end{equation}
Near the stationary point at $r=r_0$ the shift velocity according to Eq.(\ref{vSdSADM4}) is $v(r) \approx \sqrt{3} H(r-r_0)$. Then the probability of ionization in the semiclassical approximation is:
\begin{equation}
w\propto \exp\left(-2{\rm Im} \int dr\,p_r(r)\right)=\exp\left(-\frac{\pi\epsilon_0}{\sqrt{3}H}\right)\,.
\label{exponent}
\end{equation}
This process of ionization looks as thermal, $w\propto \exp(-\epsilon_0/T)$, with the temperature:
 \begin{equation}
T=\frac{\sqrt{3}H}{\pi}\equiv T_{\rm SdS}\,.
\label{Ionization0}
\end{equation}
It can be shown that the same result (\ref{Ionization0}) is valid for the general case of arbitrary $C$, i.e. for any relation between the black hole and cosmological horizons.
This suggests that the temperature $T_{\rm SdS}$ in Eq.(\ref{Ionization0}) is the universal temperature in the SdS environment, which does not depend on the mass of black hole and is fully determined by the Hubble parameter.

  \section{Probability of mass splitting}

One can apply the same procedure to calculate in the Schwarzschild-de Sitter environment the decay of a composite particle with mass $m_0$ into two particles with masses $m_1$ and $m_2$, when $m_1 + m_2 > m_0$. Such decay is also energetically forbidden
in the Minkowski spacetime, but is allowed in the dS and SdS backgrounds.
 In the simplest case, when  $m_1=m_2$, one obtains the decay probability:
\begin{equation}
w\propto \exp\left(-\frac{\pi(m_1+m_2-m_0}{\sqrt{3}H}\right)\,,
\label{exponent2}
\end{equation}
which also corresponds to thermal activation with the same temperature $T_{\rm SdS}$ in Eq.(\ref{Ionization0}), as for the process of ionization.
This is natural, since ionization is an example of decay of composite particle -- the atom.

 \section{Activation temperature vs temperatures of horizons}

The Eq.(\ref{Ionization0})  is also valid in the Nariai limit when the two horizons are merging. 
On the other hand, the merging of the two horizons can be obtained by adiabatic increase of the Newton constant $G$ at fixed $M$ and $H$. This provides the explanation, why the  temperature   $T_{\rm SdS}$ is universal: it does not change under adiabatic variation. Similar insensitivity to the adiabatic process was found for the temperature of Hawking radiation from the Reissner–Nordstr\"om black hole, which also has two horizons.\cite{Volovik2022} This temperature does not depend on the relative positions of the inner $r_-$ and outer $r_+$ horizons, and is fully determined by the mass of the black hole:
$T_{\rm H}=1/4\pi(r_+ +r_-)=1/8\pi GM$. The adiabatic parameter in this case is the fine structure constant $\alpha$, while the parameters $M$ and $G$ and the $U(1)$ charge of the black hole in terms of the electron charge are fixed.

 In the limit when the two horizons merge we can compare the activation temperature $T_{\rm SdS}$  with the Bousso-Hawking temperature in Eq.(\ref{GlobalT}), which is determined by the common horizon in this Nariai limit of the SdS spacetime.\cite{BoussoHawking1996} The activation temperature appears to be twice larger than the Bousso-Hawking temperature:
\begin{equation}
T_{\rm SdS}=2T_{\rm BH}\,.
\label{Ionization2}
\end{equation}

\section{de Sitter thermodynamics with the local temperature}
\label{dSthermo}

The relation (\ref{Ionization2}) also takes place for  the de Sitter spacetime, where the activation temperature $T_{\rm dS}$ of different processes measured by stationary observer at $r=0$ is double the Gibbons-Hawking temperature, 
$T_{\rm dS}=2T_{\rm GH}=H/\pi$.\cite{Volovik2009,Volovik2022b,Bros2008,Bros2010,Jatkar2012}  
 The discussion of the possible origins of the doubling of the Hawking temperature in different arrangements see in Ref.\cite{Volovik2022b} and references therein. 
 
The possible origin of the factor 2  comes from the de Sitter expansion considered as a particle accelerator,\cite{Maldacena2015}
 where the created particles with mass $m_0$ creates the other particles with masses $m$.
 According to Eq.(\ref{exponent2}), the decay exponent for $m_1=m_2=m$ is
 $w\propto \exp\left(-\pi(2m-m_0)/H \right)$. 
For $m_0\rightarrow 0$ this process gives  $w\propto \exp\left(-2\pi m/H \right)$. This looks as the Hawking emission of particles with mass $m$ from the de Sitter vacuum with the Gibbons-Hawking temperature $T_{\rm GH}=H/2\pi$. But in fact it corresponds to the coherent creation of two particles, $w\propto \exp\left(-\pi (2m)/H \right)$, which corresponds to the double Gibbons-Hawking temperature, see also Ref.\cite{Mottola2018}. 
  
Here we suggest another interpretation. Since the activation temperature in dS describes the local processes, which are not related to the dS cosmological horizon, one may suggest that $T=T_{\rm SdS}=H/\pi$ plays the role of the local thermodynamic temperature of the dS vacuum.  All the points in the de Sitter space are equivalent, and thus this local temperature is the same for all static observers. Then the energy density at any point of the de Sitter vacuum can be considered as the density of the thermal energy expressed in terms of this local temperature:
\begin{equation}
 \epsilon_{\rm vac}=\frac{3}{8\pi G}H^2=\frac{3\pi}{8G}T^2\,.
\label{dSEnergyDensity}
\end{equation}

In this vacuum thermodynamics, the local entropy density is
\begin{equation}
s_{\rm vac}= - \frac{\partial F}{\partial T} =\frac{3\pi}{4G}T=\frac{3}{4G}H\,,
\label{dSEntropyDensity}
\end{equation}
where the free energy density $F=\epsilon_{\rm vac} - T d\epsilon_{\rm vac}/dT$. 
The total entropy in the volume surrounded by the cosmological horizon with radius $R=1/H$ (we consider here the expanding de Sitter with $H>0$) is
\begin{equation}
S_{\rm vac}=\frac{4\pi R^3}{3} s _{\rm vac}= \frac{\pi}{GH^2}=\frac{A}{4G} \,,
\label{dSEntropy}
\end{equation}
where $A$ is the horizon area.
This corresponds to the Gibbons-Hawking entropy. But in this approach it is the thermodynamic entropy coming from the local entropy of the de Sitter quantum vacuum, rather than from the cosmological horizon.

The local temperature of the quantum vacuum suggests that in the presence of matter one nay have the quasi-equilibrium states of the Universe with different temperatures for vacuum and for matter degrees of freedom.  Such situation with the vacuum temperature $\sim H$ has been have been considered in Ref. \cite{Vergeles2023} with application to the inflationary stage.
The present temperature of the vacuum is much smaller than the temperature of matter degrees of freedom, such as the temperature of Cosmic Microwave Background (CMB) radiation,
$T_{\rm dS}\sim 10^{-30}T_{\rm CMB}$. But the entropy of the vacuum highly exceeds the entropy of matter due to large density of states in the quantum vacuum, $s_{\rm vac} \sim 10^{30} s_{\rm CMB}$. 

\section{White hole in contracting de Sitter and negative temperature}

The modified PG coordinates have been also applied to the other SdS configurations.\cite{Volovik2023} In particular, in case of the white hole in the environment of the contracting de Sitter  the shift velocity is given by Eq.(\ref{vSdSADM3}) with the opposite sign:
\begin{equation}
  v(r)=- \sqrt{C(1-C)} \,\sqrt{\frac{r+2r_0}{3rr_0^2}}\,\,(r-r_0)\,.
\label{Contracting}
\end{equation}
 Since the cosmological arrow of time changes sign, the activation temperature becomes negative, $T_{\rm SdS}=-\sqrt{3}H/\pi$. 
 
 For the contracting Universe, the negative temperature  is not so unusual. In particular, this can be obtained by extending the analytic continuation across the Big-Bang \cite{Boyle2018,Boyle2022} to the temperature axis,\cite{Volovik2019} see also Ref.\cite{Volovik2019b}. 
 For the contracting de Sitter state one can introduce the hydrodynamics written in terms of the local temperature in the approach discussed in the previous section. In the contracting state one has  $H<0$, which gives the negative temperature,  $T=H/\pi <0$. From Eq.(\ref{dSEntropyDensity}) one obtains for contracting dS vacuum the negative entropy density, $s=(3/4G)H<0$, and the negative entropy inside the cosmological horizon, $S=-A/4G$
 (the discussion of the negative entropy in different cosmological arrangements see e.g. in Ref. \cite{Volovik2022}).

\section{Conclusion}

We found that the activation temperature in Schwarzschild-de Sitter background, which is measured by the static observer, is universal.  It does not depend on the black hole mass $M$ and is fully determined by the Hubble parameter $H$ of the de Sitter expansion, although with different prefactor,
$T_{\rm SdS}=\sqrt{3}H/\pi$. This universal temperature is twice the Bousso-Hawking temperature $T_{\rm BH}=\sqrt{3}H/2\pi$, which characterizes the limit of nearly degenerate Schwarzschild–de Sitter universe, when the cosmological and black hole horizons are close to each other and radiate with almost the same temperature. The temperature $T_{\rm SdS}$ is the universal temperature, which characterizes different processes of thermal activation including the process of ionization of atoms, the decay of the composite particles in the SdS, and all the other scattering or radiation processes, which are not possible in the Minkowski spacetime. That is why it can be considered as the global temperature in the SdS spacetime.

Since these thermal processes take place away from the black hole horizon, the existence of the black hole horizon is not necessary for that. The universal temperature is only determined by the existence of the surfaces or points with zero gravity, where the static observer can measure the ionization of the atoms at rest and the other processes. This approach can be extended to the many body state in the cosmological environment, such as the static black binaries in the de Sitter spacetime.\cite{Gibbons2023} 
Also it will be interesting to consider the interconnection between the temperatures $T_{\rm SdS}$  and $T_{\rm dS}$ caused by gravity and the local temperature $T$ in the detector environment, which is similar to the case discussed in Ref.\cite{Akhmedov2022a} and which may lead to the analog of the quantum phase transition discussed in Ref. \cite{Zakharov2023} at the Unruh temperature.

 {\bf Acknowledgements}.   
 This work has been supported by Academy of Finland (grant 332964).

\end{document}